\documentclass[aps,pra,twocolumn,groupedaddress]{revtex4-1}
\usepackage{amsmath} 
\usepackage{graphicx}
\usepackage{dcolumn}
\usepackage{mathrsfs}
\usepackage{bm}
\usepackage{subfigure} 
\usepackage[colorlinks,linkcolor=blue,anchorcolor=blue,citecolor=blue,urlcolor=blue]{hyperref}

\begin{document}

\title{Entangling two microwave modes via optomechanics}
\author{Qizhi Cai$^{1}$}\thanks{E-mail: 254489627@qq.com}
\author{Jinkun Liao$^{1}$}\thanks{E-mail: jkliao@uestc.edu.cn}
\author{Qiang Zhou$^{1,2}$}
\address{$^{1}$School of optoelectronic science and engineering, University of Electronic Science and Technology of China, Chengdu, Sichuan, China}
\address{$^{2}$Institute of Fundamental and Frontier Sciences, University of Electronic Science and Technology of China, Chengdu, Sichuan, China}

\date{\today }
\begin{abstract}
We in theory proposed a hybrid system consisting of a mechanical resonator, an optical Fabry-Perot cavity and two superconducting microwave circuits to generate stationary continuous-variable quantum entanglement between two microwave modes. We show that the hybrid system can also achieve quantum entanglement of other bipartite subsystems in experimentally accessible parameter regimes, which has the potential to be useful in quantum information processing and quantum illumination radar.
\end{abstract}

\pacs{Valid PACS appear here}
\maketitle

\section{Introduction}
As a particular resource of the quantum world, entanglement plays an important role in fundamental quantum mechanics and high-speed development of quantum technologies, such as quantum internet, quantum communication, quantum sensing and quantum computer [1-5]. The realization of the above techniques must rely on the generation and distribution of entanglement between different quantum systems, including atoms, spins, photons, ions, phonons and superconducting circuits [6-12].

Owing to the strong coupling with external microwave field, superconducting circuit can control, store and read the quantum information robustly with good scalability, which makes it a promising candidate for the physical implementation of quantum computer. Moreover, entangled microwaves can be used in quantum-enhanced radar schemes, namely quantum illumination radar protocol, which utilizes quantum resources to detect targets with low-reflectivity hidden in a strong thermal noise bath, in the case where quantum entanglement is destroyed, its detection ability can still maintain quantum superiority, namely exceeding the classical counterpart protocols [13-15]. Nowadays, the realization of the prototype quantum radar has begun to land the ground in the microwave band, and it shows robustness to the ambient background noise and loss, which means it may have broad application prospects [16-17]. The above series of requirements and others make the generation of entangled microwave and the distribution of its entanglement become the research hotspots, and various theoretical schemes and physical experiments to realize entanglement between microwaves have emerged one after another in recent years [18-21].

In this work, we propose a theoretical scheme to achieve the entanglement of two microwave modes mediated by optomechanics. This hybrid quantum system consists of a mechanical resonator, two superconducting circuits and an Fabry-Perot optical cavity. It could be realized with the help of lumped-element superconducting circuit where a free-standing mechanical membrane acting as drum-head capacitor, which is capacitively coupled to two microwave cavities simultaneously (see Fig.1). The drum-head capacitor could be optically coated to form a Fabry-Perot optical cavity with the other side of the micromirror. Therefore, the vibrating drum-head mechanical membrane could capacitively interface with two microwave cavities at the same time while the optical intracavity mode is coupled with the drum-head by the radiation pressure.

This quadripartite optoelectromechanical system has the potential to be used in quantum computer and quantum communication networks and quantum illumination radar systems. The optical intracavity mode is not susceptible to thermal noise, so it could be used as a flying qubit to interface with other distant node via fibers. The quantum entanglement between microwaves, light and microwave could be distributed to other network nodes through the entanglement swapping protocol [22-25]. In addition, this quadripartite system can be fabricated on chips [26-28], which proposes a new approach to realize the integrated microwave source for quantum illumination radar systems.

This paper is organized as follows. Sec. II shows the basic physical model. The quantum Langevin equations with their linearization for describing the dynamics of the hybrid system are discussed in Sec. III. In Sec. IV, we will derive the correlated matrix of the quantum fluctuation of the system to obtain the logarithmic negativity, which is considered as the entanglement measure in this work. In Sec. V, we study the effects of different parameters on the quantum entanglement between microwave modes, and also analyze the entanglement between other bipartite subsystems like light-microwave subsystem and so on, while Sec. VI is for conclusion.

\section{Model}
The proposed scheme is depicted in Fig.1. On the one hand, the mechanical resonator with resonate frequency ${\omega _m}$ is capacitively coupled to two superconducting circuits simultaneously, with the two superconducting circuits with resonate frequency ${\omega _{w1}}$, ${\omega _{w2}}$, respectively. On the other hand, the mechanical resonator is coupled to a Fabry-Perot optical cavity with resonate frequency ${\omega _c}$. The optical and microwave cavities are driven at the frequencies ${\omega _{0c}} = {\omega _c} - {\Delta _{0c}}$, ${\omega _{0w1}} = {\omega _{w1}} - {\Delta _{0w1}}$ and ${\omega _{0w2}} = {\omega _{w2}} - {\Delta _{0w2}}$, respectively. The effective Hamiltonian for the quadripartite system is [29-31]
\begin{figure}
	\centering
	\includegraphics[width=1\linewidth]{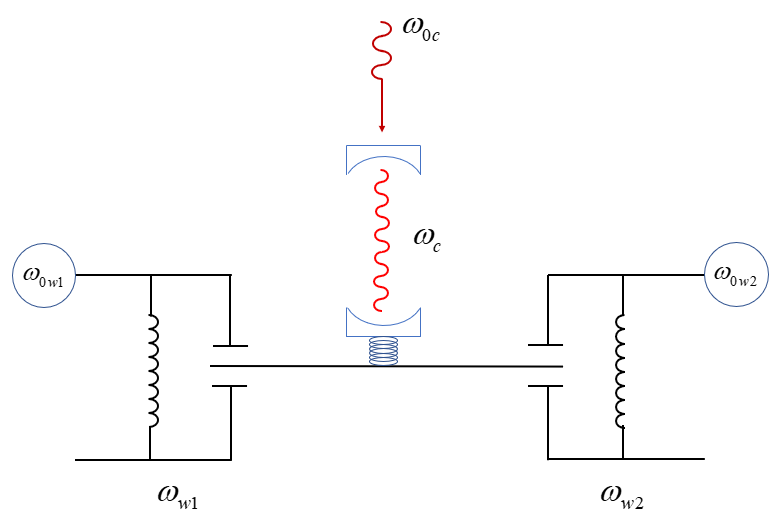}
	\caption{Simple schematic diagram of the hybrid system}
	\label{fig:Fig1.png}
\end{figure}
\begin{eqnarray}
\begin{aligned}
H =& \frac{{\hat p_x^2}}{{2m}} + \frac{{m\omega _m^2{{\hat x}^2}}}{2} + \frac{{\hat \Phi _1^2}}{{2{L_1}}} + \frac{{\hat Q_1^2}}{{2[{C_1} + {C_{d1}}(x)]}} - {e_1}(t){{\hat Q}_1}\\
&+ \frac{{\hat \Phi _2^2}}{{2{L_2}}} + \frac{{\hat Q_2^2}}{{2[{C_2} + {C_{d2}}(x)]}} - {e_2}(t){{\hat Q}_2}\\
&+ \hbar {\omega _c}{{\hat a}^\dag }\hat a - \hbar {G_{0c}}{{\hat a}^\dag }\hat a\hat x + i\hbar {E_c}({{\hat a}^\dag }{e^{ - i{\omega _{0c}}t}} - \hat a{e^{i{\omega _{0c}}t}}).
\end{aligned}
\end{eqnarray}
where $(\hat x,\hat p_x)$ are the canonical position and momentum of a mechanical resonator, $(\hat \Phi_j,\hat Q_j)$ are the canonical coordinates for the two microwave cavities ($j=1,2$ is a notation for the two superconducting microwave ciucuits), indicating the flux through equivalent inductors $L_j$ and the charge on equivalent capacitors $C_j$, respectively. $(\hat a,{\hat a}^\dag)$ is the annihilation and creation operator of the optical cavity mode, which satisfies $[\hat a,{\hat a^\dag }] = 1$, and ${E_c} = \sqrt {{{2{P_c}{\kappa _c}} \mathord{\left/
			{\vphantom {{2{P_c}{\kappa _c}} {\hbar {\omega _{0c}}}}} \right.
			\kern-\nulldelimiterspace} {\hbar {\omega _{0c}}}}} $ is related to the laser input, where $P_c$ is the power of the input laser and $\kappa _c$ describes the damping rate of the optical cavity. ${G_{0c}} = ({{{\omega _c}} \mathord{\left/
		{\vphantom {{{\omega _c}} {L)}}} \right.
	\kern-\nulldelimiterspace} {l)}}\sqrt {{\hbar  \mathord{\left/
	{\vphantom {\hbar  {m{\omega _m}}}} \right.
	\kern-\nulldelimiterspace} {m{\omega _m}}}} $ is the coupling between optical cavity and mechanical resonator, with $l$ the length of the optical Fabry-Perot cavity and $m$ the effective mass of mechainical mode. The coherent drving of the superconducting microwave cativies with damping rates $\kappa _{wj}$ is given by the electric potential ${e_j}(t) =  - i\sqrt {2\hbar {\omega _{wj}}{L_j}} {E_{wj}}({e^{i{\omega _{0wj}}t}} - {e^{ - i{\omega _{0wj}}t}})$, where ${E_{wj}} = \sqrt {{{2{P_{wj}}{\kappa _{wj}}} \mathord{\left/
	{\vphantom {{2{P_{wj}}{\kappa _{wj}}} {\hbar {\omega _{0wj}}}}} \right.
	\kern-\nulldelimiterspace} {\hbar {\omega _{0wj}}}}} $. The reason of the coupling between mechanical resonator and microwave cavities is that the capacities are functions of the resonator displacement, namely $C_{dj} (x)$. We expand these functions around their equilibrium positions $d_j$ for two superconducting microwave cavities, then we have ${C_{dj}}(x) = {C_{dj}}[1 - {{{x_j}(t)} \mathord{\left/
{\vphantom {{{x_j}(t)} {{d_j}}}} \right.
\kern-\nulldelimiterspace} {{d_j}}}]$. Expanding the capacitive energy as a Taylor series, we find to first order,
\begin{equation}
\begin{split}
\frac{{\hat Q_j^2}}{{2[{C_j} + {C_{dj}}({x_j})]}} = \frac{{\hat Q_j^2}}{{2{C_{\Sigma j}}}} - \frac{{{\mu _j}}}{{2{d_j}{C_{\Sigma j}}}}{\hat x_j}(t)\hat Q_j^2.
\end{split}
\end{equation}
where ${C_{\Sigma j}} = {C_j} + {C_{dj}}$ and ${\mu _j} = {{{C_{dj}}} \mathord{\left/
		{\vphantom {{{C_{dj}}} {{C_{\Sigma j}}}}} \right.
		\kern-\nulldelimiterspace} {{C_{\Sigma j}}}}$. In this manner, the Hamiltonian of Eq.(1) can be reshaped in the terms of the annihilation and creation operators of the superconducting microwave cavities' field $(\hat b_j,{\hat b_j}^\dag)$ and the dimensionless position and momentum operators of the mechanical resonator $(\hat q,\hat p)$, which satisfy $[\hat b_j,{\hat b_j}^\dag] = 1$ and $[\hat q,\hat p] = i$, as
\begin{equation}
\begin{split}
H =& \hbar {\omega _c}{{\hat a}^\dag }\hat a + \hbar {\omega _{w1}}\hat b_1^\dag {{\hat b}_1} + \hbar {\omega _{w2}}\hat b_2^\dag {{\hat b}_2}+ \frac{{\hbar {\omega _m}}}{2}({{\hat p}^2} 
+ {{\hat q}^2})\\ 
&- \frac{{\hbar {G_{0w1}}}}{2}\hat q{({{\hat b}_1} + \hat b_1^\dag )^2} - \frac{{\hbar {G_{0w2}}}}{2}\hat q{({{\hat b}_2} + \hat b_2^\dag )^2}\\
&- \hbar {G_{0c}}\hat q{{\hat a}^\dag }\hat a - i\hbar {E_{w1}}({e^{i{\omega _{0w1}}t}} - {e^{ - i{\omega _{0w1}}t}})({{\hat b}_1} + \hat b_1^\dag )\\
&- i\hbar {E_{w2}}({e^{i{\omega _{0w2}}t}} - {e^{ - i{\omega _{0w2}}t}})({{\hat b}_2} + \hat b_2^\dag )\\
&+ i\hbar {E_c}({{\hat a}^\dag }{e^{ - i{\omega _{0c}}t}} - \hat a{e^{i{\omega _{0c}}t}}).
\end{split}
\end{equation}
where
\begin{eqnarray}
\begin{aligned}
{\hat b_j} = \sqrt {\frac{{{\omega _{wj}}{L_j}}}{{2\hbar }}} {\hat Q_j} + \frac{i}{{\sqrt {2\hbar {\omega _{wj}}{L_j}} }}{\hat \Phi _j},
\end{aligned}
\end{eqnarray}
\begin{eqnarray}
\begin{aligned}
\hat q = \sqrt {\frac{{m{\omega _m}}}{\hbar }} \hat x,\hat p = \frac{{{{\hat p}_x}}}{{\sqrt {\hbar m{\omega _m}} }},
\end{aligned}
\end{eqnarray}
\begin{eqnarray}
\begin{aligned}
{G_{0wj}} = \frac{{{\mu _j}{\omega _{wj}}}}{{2{d_j}}}\sqrt {\frac{\hbar }{{m{\omega _m}}}}. 
\end{aligned}
\end{eqnarray}
Typically, ${\omega _m} <  < {\omega _c},{\omega _{wj}}$ since the mechanical resonant frequency is approximately 10 MHz while the optical and microwave cavity resonant frequency are the order of 10 GHz and 100 THz, respectively. Therefore, it is convenient to move into the interaction picture with respect to ${H_0} = \hbar {\omega _{0c}}{\hat a^\dag }\hat a + \hbar {\omega _{0w1}}\hat b_1^\dag {\hat b_1} + \hbar {\omega _{0w2}}\hat b_2^\dag {\hat b_2}$, and neglect fast oscillating terms at $ \pm 2{\omega _{0c}}, \pm 2{\omega _{0wj}}$. In this case, the corresponding Hamiltonian of the system becomes
\begin{eqnarray}
\begin{aligned}
H =& \hbar {\Delta _{0c}}{{\hat a}^\dag }\hat a + \hbar {\Delta _{0w1}}\hat b_1^\dag {{\hat b}_1} + \hbar {\Delta _{0w2}}\hat b_2^\dag {{\hat b}_2}\\
&+ \frac{{\hbar {\omega _m}}}{2}({{\hat p}^2} + {{\hat q}^2}) - \hbar {G_{0w1}}\hat q\hat b_1^\dag {{\hat b}_1} - \hbar {G_{0w2}}\hat q\hat b_2^\dag {{\hat b}_2}\\
&- \hbar {G_{0c}}\hat q{{\hat a}^\dag }\hat a - i\hbar {E_{w1}}({{\hat b}_1} - \hat b_1^\dag )\\
&- i\hbar {E_{w2}}({{\hat b}_2} - \hat b_2^\dag ) + i\hbar {E_c}({{\hat a}^\dag } - \hat a).
\end{aligned}
\end{eqnarray}

\section{Quantum Langivin equations and their linearization}
However, the system would inevitably interact with the surrounding environment, bringing damping and noise to affect each mode. We can decribe them with the help of quantum Langevin equation where the Heisenberg equations for the system operators is revised by inserting the corresponding damping and noise terms. In this way, the nonlinear quantum Langevin equations of the system is given by
\begin{eqnarray}
\begin{aligned}
\dot{\hat{q}} = {\omega _m}\hat p, 
\end{aligned}
\end{eqnarray}
\begin{eqnarray}
\begin{aligned}
\dot{\hat{p}} =&  - {\omega _m}\hat q - {\kappa _m}\hat p + {G_{0c}}{\hat a^\dag }\hat a\\ 
&+ {G_{0w1}}\hat b_1^\dag {\hat b_1} + {G_{0w2}}\hat b_2^\dag {\hat b_2} + \xi, 
\end{aligned}
\end{eqnarray}
\begin{eqnarray}
\begin{aligned}
\dot{\hat{a}} =  - (i{\Delta _{0c}} + {\kappa _c})\hat a + i{G_{0c}}\hat q\hat a + {E_c} + \sqrt {2{\kappa _c}} {\hat a_{in}}, 
\end{aligned}
\end{eqnarray}
\begin{equation}
\begin{split}
\dot{\hat{b}}_1 = - (i{\Delta _{0w1}} + {\kappa _{w1}}){\hat b_1} + i{G_{0w1}}\hat q{\hat b_1} + {E_{w1}} + \sqrt {2{\kappa _{w1}}} {\hat b_{in,1}}, 
\end{split}
\end{equation}
\begin{equation}
\begin{split}
\dot{\hat{b}}_2 = - (i{\Delta _{0w2}} + {\kappa _{w2}}){\hat b_2} + i{G_{0w2}}\hat q{\hat b_2} + {E_{w2}} + \sqrt {2{\kappa _{w2}}} {\hat b_{in,2}}, 
\end{split}
\end{equation}
where $\hat a_{in}$ and $\hat b_{in,j}$ are the input noise terms of optical and microwave modes, respectively, which could be considered as zero-mean Gaussian processes, satisfying the following correlation [32]
\begin{eqnarray}
\begin{aligned}
\left\langle {{{\hat a}_{in}}(t)\hat a_{in}^\dag (t')} \right\rangle  = [N({\omega _c}) + 1]\delta (t - t'), 
\end{aligned}
\end{eqnarray}
\begin{eqnarray}
\begin{aligned}
\left\langle {\hat a_{in}^\dag (t){{\hat a}_{in}}(t')} \right\rangle  = N({\omega _c})\delta (t - t'), 
\end{aligned}
\end{eqnarray}
\begin{eqnarray}
\begin{aligned}
\left\langle {{{\hat b}_{in,j}}(t)\hat b_{in,j}^\dag (t')} \right\rangle  = [N({\omega _{wj}}) + 1]\delta (t - t'), 
\end{aligned}
\end{eqnarray}
\begin{eqnarray}
\begin{aligned}
\left\langle {\hat b_{in,j}^\dag (t){{\hat b}_{in,j}}(t')} \right\rangle  = N({\omega _{wj}})\delta (t - t'), 
\end{aligned}
\end{eqnarray}
in which $N({\omega _c}) = {1 \mathord{\left/
		{\vphantom {1 {[\exp ({{\hbar {\omega _c}} \mathord{\left/
							{\vphantom {{\hbar {\omega _c}} {{k_B}T) - 1]}}} \right.
							\kern-\nulldelimiterspace} {{k_B}T) - 1]}}}}} \right.
		\kern-\nulldelimiterspace} {[\exp ({{\hbar {\omega _c}} \mathord{\left/
				{\vphantom {{\hbar {\omega _c}} {{k_B}T) - 1]}}} \right.
				\kern-\nulldelimiterspace} {{k_B}T) - 1]}}}}$ and $N({\omega _{wj}}) = {1 \mathord{\left/
			{\vphantom {1 {[\exp ({{\hbar {\omega _{wj}}} \mathord{\left/
						{\vphantom {{\hbar {\omega _{wj}}} {{k_B}T) - 1]}}} \right.
						\kern-\nulldelimiterspace} {{k_B}T) - 1]}}}}} \right.
		\kern-\nulldelimiterspace} {[\exp ({{\hbar {\omega _{wj}}} \mathord{\left/
		{\vphantom {{\hbar {\omega _{wj}}} {{k_B}T) - 1]}}} \right.
		\kern-\nulldelimiterspace} {{k_B}T) - 1]}}}}$ are the mean thermal excitatiion numbers of optical and microwaves fields, respectively, where $k_B$ is the Boltzmann constant and $T$ is the temperature of the surrounding environment. Since ${{\hbar {\omega _c}} \mathord{\left/
	{\vphantom {{\hbar {\omega _c}} {{k_B}T}}} \right.
\kern-\nulldelimiterspace} {{k_B}T}} >  > 1$ at optical frequencies, we can safely assume that $N({\omega _c}) \to 0$, while $N({\omega _{wj}})$ cannot be ignored even at quite low temperatures. What's more, in Eq.(9), $\kappa _{m}$ is the mechanical damping rate and $\xi (t)$ is the quantum Brownian noise acting on the mechanical resonator, with the correlation function [33]
\begin{equation}
\begin{split}
\left\langle {\xi (t)\xi (t')} \right\rangle  = \frac{{{\kappa _m}}}{{{\omega _m}}}\int {\frac{{d\omega }}{{2\pi }}} {e^{ - i\omega (t - t')}}\omega [\coth (\frac{{\hbar \omega }}{{2{k_B}T}}) + 1].
\end{split}
\end{equation}
$\xi (t)$ is not delta-correlated obviously and therefore does not describe a Markovian process. But the mechanical quantum effects are achieved only for a high mechanical quality factor (${Q_m} = {{{\omega _m}} \mathord{\left/
		{\vphantom {{{\omega _m}} {{\kappa _m}}}} \right.
		\kern-\nulldelimiterspace} {{\kappa _m}}} >  > 1$), in this limit, $\xi (t)$ becomes
delta-correlated [34], i.e., ${{\left\langle {\xi (t)\xi (t') + \xi (t')\xi (t)} \right\rangle } \mathord{\left/
		{\vphantom {{\left\langle {\xi (t)\xi (t') + \xi (t')\xi (t)} \right\rangle } 2}} \right.
		\kern-\nulldelimiterspace} 2} \approx {\kappa _m}(2{\bar n_m} + 1)\delta (t - t')$, where ${\bar n_m} = {1 \mathord{\left/
		{\vphantom {1 {[\exp ({{\hbar {\omega _m}} \mathord{\left/
							{\vphantom {{\hbar {\omega _m}} {{k_B}T) - 1]}}} \right.
							\kern-\nulldelimiterspace} {{k_B}T) - 1]}}}}} \right.
		\kern-\nulldelimiterspace} {[\exp ({{\hbar {\omega _m}} \mathord{\left/
				{\vphantom {{\hbar {\omega _m}} {{k_B}T) - 1]}}} \right.
				\kern-\nulldelimiterspace} {{k_B}T) - 1]}}}}$ is mean thermal excitation number of the mechanical resonator like microwave and optical counterparts mentioned above.

In order to achieve stationary and robust entanglement in continuous-variable systems, we would like to choose an working point where the cavities are intensely driven. In this way, the intracavity fields are strong enough to do the following approximation: $\hat a=\alpha _s+\delta \hat a$, $\hat b_j=\beta _{js}+\delta \hat b_j$, $\hat p=p _s+\delta \hat p$ and $\hat q=q _s+\delta \hat q$, in which $\alpha _s$, $\beta _{js}$, $p _s$, $q _s$ and $\delta \hat a$, $\delta \hat b_j$, $\delta \hat p$, $\delta \hat q$ are the fixed semiclassical points and their quantum fluctuation around the semiclassical points of quadripartite system, respectively. Then we insert the above approximation into Eq.(8-12) and let the derivatives to zero, getting the fixed points 
\begin{equation}
\begin{split}
{p_s} = 0, 
\end{split}
\end{equation}
\begin{eqnarray}
\begin{aligned}
{q_s} = \frac{{{G_{0c}}{{\left| {{\alpha _s}} \right|}^2} + {G_{0w1}}{{\left| {{\beta _{1s}}} \right|}^2} + {G_{0w2}}{{\left| {{\beta _{2s}}} \right|}^2}}}{{{\omega _m}}}, 
\end{aligned}
\end{eqnarray}
\begin{eqnarray}
\begin{aligned}
{\alpha _s} = \frac{{{E_c}}}{{{\kappa _c} + i{\Delta _c}}}, 
\end{aligned}
\end{eqnarray}
\begin{eqnarray}
\begin{aligned}
{\beta _{js}} = \frac{{{E_{wj}}}}{{{\kappa _{wj}} + i{\Delta _{wj}}}}, 
\end{aligned}
\end{eqnarray}
where ${\Delta _c} = {\Delta _{0c}} - {G_{0c}}{q_s}$ and ${\Delta _{wj}} = {\Delta _{0wj}} - {G_{0wj}}{q_s}$ are the effective detunning of the optical and microwaves fields, respectively. As mentioned above, the optical and microwaves intracavity fields are intensely driven, namely $\left| {{\alpha _s}} \right| >  > 1$ and $\left| {{\beta _{sj}}} \right| >  > 1$, thus, we can securely linearized Eq.(8-12) around the fixed semiclassical points and get the following linear quantum Langevin equations for the quantum fluctuations of the hybrid system:

\begin{eqnarray}
\begin{aligned}
\delta\dot{\hat{q}} = {\omega _m}\delta \hat p, 
\end{aligned}
\end{eqnarray}
\begin{equation}
\begin{split}
\delta\dot{\hat{p}} =& - {\omega _m}\delta \hat q - {\kappa _m}\delta \hat p + {G_{0c}}{\alpha _s}(\delta {\hat a^\dag } + \delta \hat a)\\
&+ {G_{0w1}}{\beta _{1s}}(\delta {\hat b_1}^\dag  + \delta {\hat b_1}) + {G_{0w2}}{\beta _{2s}}(\delta {\hat b_2}^\dag  + \delta {\hat b_2}) + \xi, 
\end{split}
\end{equation}
\begin{eqnarray}
\begin{aligned}
\delta\dot{\hat{a}} =  - ({\kappa _c} + i{\Delta _c})\delta \hat a + i{G_{0c}}{\alpha _s}\delta \hat q + \sqrt {2{\kappa _c}} {\hat a_{in}}, 
\end{aligned}
\end{eqnarray}
\begin{equation}
\begin{split}
\delta\dot{\hat{b}}_1 =  - ({\kappa _{w1}} + i{\Delta _{w1}})\delta {\hat b_1} + i{G_{0w1}}{\beta _{1s}}\delta \hat q + \sqrt {2{\kappa _{w1}}} {\hat b_{in,1}}, 
\end{split}
\end{equation}
\begin{equation}
\begin{split}
\delta\dot{\hat{b}}_2 =  - ({\kappa _{w2}} + i{\Delta _{w2}})\delta {\hat b_2} + i{G_{0w2}}{\beta _{2s}}\delta \hat q + \sqrt {2{\kappa _{w2}}} {\hat b_{in,2}}, 
\end{split}
\end{equation}
where we have chosen the appropriate reference phases for the three input optical and microwaves fields so that $\alpha_s$ and $\beta_{js}$ can be taken as real and positive.

\section{Correlated matrix of quantum fluctuations}
The continuous-variable entanglement between the two microwave cavities is generated by the microwave cavities' interaction mediated by the optomechanical bipartite system, in other word, the quantum correlations between the quadratures of three intracavity fields and the position and momentum of the mechanical resonator. Based on this reason, it is favourable to introduce the fluctuation quadratures $\delta {\hat X_c} = {{(\delta \hat a + \delta {{\hat a}^\dag })} \mathord{\left/
		{\vphantom {{(\delta \hat a + \delta {{\hat a}^\dag })} {\sqrt 2 }}} \right.
		\kern-\nulldelimiterspace} {\sqrt 2 }}$ and $\delta {\hat Y_c} = {{(\delta \hat a - \delta {{\hat a}^\dag })} \mathord{\left/
		{\vphantom {{(\delta \hat a - \delta {{\hat a}^\dag })} {i\sqrt 2 }}} \right.
		\kern-\nulldelimiterspace} {i\sqrt 2 }}$ for the optical field, $\delta {\hat X_{wj}} = {{(\delta {{\hat b}_j} + \delta \hat b_j^\dag )} \mathord{\left/
		{\vphantom {{(\delta {{\hat b}_j} + \delta \hat b_j^\dag )} {\sqrt 2 }}} \right.
		\kern-\nulldelimiterspace} {\sqrt 2 }}$ and $\delta {\hat Y_{wj}} = {{(\delta {{\hat b}_j} - \delta \hat b_j^\dag )} \mathord{\left/
		{\vphantom {{(\delta {{\hat b}_j} - \delta \hat b_j^\dag )} {i\sqrt 2 }}} \right.
		\kern-\nulldelimiterspace} {i\sqrt 2 }}$ for the two microwave fields. The corresponding input noise operators are ${\hat X_{c,in}} = {{(\delta {{\hat a}_{in}} + \delta \hat a_{in}^\dag )} \mathord{\left/
		{\vphantom {{(\delta {{\hat a}_{in}} + \delta \hat a_{in}^\dag )} {\sqrt 2 }}} \right.
		\kern-\nulldelimiterspace} {\sqrt 2 }}$, ${\hat Y_{c,in}} = {{(\delta {{\hat a}_{in}} - \delta \hat a_{in}^\dag )} \mathord{\left/
		{\vphantom {{(\delta {{\hat a}_{in}} - \delta \hat a_{in}^\dag )} {i\sqrt 2 }}} \right.
		\kern-\nulldelimiterspace} {i\sqrt 2 }}$, ${\hat X_{wj,in}} = {{(\delta {{\hat b}_{in,j}} + \delta \hat b_{in,j}^\dag )} \mathord{\left/
		{\vphantom {{(\delta {{\hat b}_{in,j}} + \delta \hat b_{in,j}^\dag )} {\sqrt 2 }}} \right.
		\kern-\nulldelimiterspace} {\sqrt 2 }}$ and ${\hat Y_{wj,in}} = {{(\delta {{\hat b}_{in,j}} - \delta \hat b_{in,j}^\dag )} \mathord{\left/
		{\vphantom {{(\delta {{\hat b}_{in,j}} - \delta \hat b_{in,j}^\dag )} {i\sqrt 2 }}} \right.
		\kern-\nulldelimiterspace} {i\sqrt 2 }}$. In this way, from Eqs.(22-26), the linearized quantum Langevin equations become 
\begin{eqnarray}
\begin{aligned}
\delta\dot{\hat{q}} = {\omega _m}\delta \hat p, 
\end{aligned}
\end{eqnarray}
\begin{eqnarray}
\begin{aligned}
\delta\dot{\hat{p}} = &- {\omega _m}\delta \hat q - {\kappa _m}\delta \hat p + {G_c}\delta {\hat X_c}\\
&+ {G_{w1}}\delta {\hat X_{w1}} + {G_{w2}}\delta {\hat X_{w2}} + \xi, 
\end{aligned}
\end{eqnarray}
\begin{eqnarray}
\begin{aligned}
\delta\dot{\hat{X}}_c =  - {\kappa _c}\delta {\hat X_c} + {\Delta _c}\delta {\hat Y_c} + \sqrt {2{\kappa _c}} {\hat X_{c,in}}, 
\end{aligned}
\end{eqnarray}
\begin{eqnarray}
\begin{aligned}
\delta\dot{\hat{Y}}_c =  - {\kappa _c}\delta {\hat Y_c} - {\Delta _c}\delta {\hat X_c} + {G_c}\delta \hat q + \sqrt {2{\kappa _c}} {\hat Y_{c,in}}, 
\end{aligned}
\end{eqnarray}
\begin{equation}
\begin{split}
\delta\dot{\hat{X}}_{w1} =  - {\kappa _{w1}}\delta {\hat X_{w1}} + {\Delta _{w1}}\delta {\hat Y_{w1}} + \sqrt {2{\kappa _{w1}}} {\hat X_{w1,in}}, 
\end{split}
\end{equation}
\begin{equation}
\begin{split}
\delta\dot{\hat{Y}}_{w1} =  - {\kappa _{w1}}\delta {\hat Y_{w1}} - {\Delta _{w1}}\delta {\hat X_{w1}} + {G_{w1}}\delta \hat q + \sqrt {2{\kappa _{w1}}} {\hat Y_{w1,in}}, 
\end{split}
\end{equation}
\begin{equation}
\begin{split}
\delta\dot{\hat{X}}_{w2} =  - {\kappa _{w2}}\delta {\hat X_{w2}} + {\Delta _{w2}}\delta {\hat Y_{w2}} + \sqrt {2{\kappa _{w2}}} {\hat X_{w2,in}}, 
\end{split}
\end{equation}
\begin{equation}
\begin{split}
\delta\dot{\hat{Y}}_{w2} =  - {\kappa _{w2}}\delta {\hat Y_{w2}} - {\Delta _{w2}}\delta {\hat X_{w2}} + {G_{w2}}\delta \hat q + \sqrt {2{\kappa _{w2}}} {\hat Y_{w2,in}}, 
\end{split}
\end{equation}
where
\begin{equation}
\begin{split}
{G_{wj}} = \sqrt 2 {G_{0wj}}{\beta _{sj}} = \frac{{{\mu _j}{\omega _{wj}}}}{{{d_j}}}\sqrt {\frac{{{P_{wj}}{\kappa _{wj}}}}{{m{\omega _m}{\omega _{0wj}}(\kappa _{wj}^2 + \Delta _{wj}^2)}}},
\end{split}
\end{equation}
\begin{equation}
\begin{split}
{G_c} = \sqrt 2 {G_{0c}}{\alpha _s} = \frac{{2{\omega _c}}}{L}\sqrt {\frac{{{P_c}{\kappa _c}}}{{m{\omega _m}{\omega _{0c}}(\kappa _c^2 + \Delta _c^2)}}},
\end{split}
\end{equation}
are the effective the electromechanical and optomechanical couplings, respectively. Equations (27-34) can be rewritten as a matrix form
\begin{equation}
\begin{split}
\dot v(t) = Av(t) + n(t),
\end{split}
\end{equation}
\begin{widetext}
in which $v(t) = {(\delta \hat q(t),\delta \hat p(t),\delta {\hat X_c}(t),\delta {\hat Y_c}(t),\delta {\hat X_{w1}}(t),\delta {\hat Y_{w1}}(t),\delta {\hat X_{w2}}(t),\delta {\hat Y_{w2}}(t))^T}$
(the notition $T$ means matrix transport), $n(t) = {(0,\xi (t),\sqrt {2{\kappa _c}} \delta {\hat X_{c,in}},\sqrt {2{\kappa _c}} \delta {\hat Y_{c,in}},\sqrt {2{\kappa _{w1}}} \delta {\hat X_{w1,in}},\sqrt {2{\kappa _{w1}}} \delta {\hat Y_{w1,in}},\sqrt {2{\kappa _{w2}}} \delta {\hat X_{w2,in}},\sqrt {2{\kappa _{w2}}} \delta {\hat Y_{w2,in}})^T}$ and	 
\begin{eqnarray}
A = \left( {\begin{array}{*{20}{c}}
	0&{{\omega _m}}&0&0&0&0&0&0\\
	{ - {\omega _m}}&{ - {\kappa _m}}&{{G_c}}&0&{{G_{w1}}}&0&{{G_{w2}}}&0\\
	0&0&{ - {\kappa _c}}&{{\Delta _c}}&0&0&0&0\\
	{{G_c}}&0&{ - {\Delta _c}}&{ - {\kappa _c}}&0&0&0&0\\
	0&0&0&0&{ - {\kappa _{w1}}}&{{\Delta _{w1}}}&0&0\\
	{{G_{w1}}}&0&0&0&{ - {\Delta _{w1}}}&{ - {\kappa _{w1}}}&0&0\\
	0&0&0&0&0&0&{ - {\kappa _{w2}}}&{{\Delta _{w2}}}\\
	{{G_{w2}}}&0&0&0&0&0&{ - {\Delta _{w2}}}&{ - {\kappa _{w2}}}
	\end{array}} \right),
\end{eqnarray}
the solution of Eq.(37) is
\begin{equation}
\begin{split}
v(t) = M(t)v(0) + \int_0^t {dsM(s)n(s)},
\end{split}
\end{equation}
where $M(t) = \exp (At)$.
\end{widetext}
In order to test the stability of the hybrid system, we use the Routh-Hurwitz criterion [35]. If the real part of all eigenvalues of matrix $A$ is negative, then we can assume that the system is stable and will approach a steady state. However, the explicit expression is too cumbersome, so we omit it here. With no special statement (Fig.2 has), the parameters used in the numerical simulation for logarithmic negativity, i.e. entanglement, of the hybrid system will satisfy the Routh-Hurwitz criterion, that is, our numerical simulation is carried out under the condition that the system is always stable.

As we know, the quantum noise terms in Eq.(37) are zero-mean Gaussian and the dynamics are linear so that the steady state of the quantum fluctuations is a continuous-varaible quadripartite Gaussian state, entirely characterized by the $8 \times 8$ correlation matrix. This matrix has components ${V_{ij}} = {{\left\langle {{u_i}(\infty ){u_j}(\infty ) + {u_j}(\infty ){u_i}(\infty )} \right\rangle } \mathord{\left/
		{\vphantom {{\left\langle {{u_k}(\infty ){u_l}(\infty ) + {u_l}(\infty ){u_k}(\infty )} \right\rangle } 2}} \right.
		\kern-\nulldelimiterspace} 2}$. When the system is stale, by using Eq.(39), we can get
\begin{equation}
\begin{split}
{V_{ij}} = \sum\limits_{k,l} {\int_0^\infty  {ds} } \int_0^\infty  {ds'} M{(s)_{ik}}M{(s')_{jl}}\Phi {(s - s')_{kl}},
\end{split}
\end{equation}
in which ${\Phi _{kl}}(s - s') = {{\left\langle {{n_k}(s){n_l}(s') + {n_l}(s'){n_k}(s)} \right\rangle } \mathord{\left/
		{\vphantom {{\left\langle {{n_k}(s){n_l}(s') + {n_l}(s'){n_k}(s)} \right\rangle } 2}} \right.
		\kern-\nulldelimiterspace} 2}$ is the matrix of stationary noise correlation functions. Here ${\Phi _{kl}}(s - s') = {D_{kl}}\delta (s - s')$, where
$D = Diag[0,{\kappa _m}(2{\bar n_m} + 1),{\kappa _c},{\kappa _c},{\kappa _{w1}}(2N({\omega _{w1}}) + 1),{\kappa _{w1}}(2N({\omega _{w1}}) + 1),{\kappa _{w2}}(2N({\omega _{w2}}) + 1),{\kappa _{w2}}(2N({\omega _{w2}}) + 1)]$ is a 8-dim diagonal matrix, in this way, Eq.(40) becomes
\begin{equation}
\begin{split}
V = \int_0^\infty  {dsM(s)D{M^T}(s)}.
\end{split}
\end{equation}	
Under the condition that the system is stable $M(\infty ) = 0$, by Lyapunov's first theorem [33], Eq.(41) is equivalent to
\begin{equation}
\begin{split}
AV + V{A^T} =  - D.
\end{split}
\end{equation}
By solving the Eq.(42), the $8\times 8$ matrix $V$ can be obtained. Then, we can calculate the entanglement of the interested bipartite systems, like two superconducting microwave circuits, light-microwave subsystems and so on, by tracing out the uninterested rows and columes of $V$. After this kind of operation, the induced correlation matrix of the interested bipartite system is a $4\times 4$ matrix as follows
\begin{equation}
\begin{split}
{V_{bi}} = \left( {\begin{array}{*{20}{c}}
	{{V_1}}&{{V_3}}\\
	{V_3^T}&{{V_2}}
	\end{array}} \right).
\end{split}
\end{equation}
Furthermore, we use the logarithmic negativity to quantify the entanglement of the interested bipartite system [37-38]
\begin{eqnarray}
E_{N}=max[0,-ln2\eta ^{-}],
\end{eqnarray}
where ${\eta ^ - } \equiv {2^{{{ - 1} \mathord{\left/
				{\vphantom {{ - 1} 2}} \right.
				\kern-\nulldelimiterspace} 2}}}{[\Sigma ({V_{bi}}) - \sqrt {\Sigma {{({V_{bi}})}^2} - 4\det {V_{bi}}} ]^{{1 \mathord{\left/
				{\vphantom {1 2}} \right.
				\kern-\nulldelimiterspace} 2}}}$ and $\Sigma ({V_{bi}}) \equiv $det${V_1} + $det${V_2} - 2$det${V_3}$. The corresponding numerical simulation is discussed in the next section.

\section{Results}
For the purpose of the hybrid system described herein can be realized by existing experimental approaches, the following simulation parameters are based on the reference [39-40] and their feasible extension. 
\begin{figure}
	\centering
	\subfigure[ The microwave detuning range is from -0.8 to 0 $\Delta_{w}/\omega_m$.]{
		\label{fig:Fig2(a).png} 
		\includegraphics[width=1\linewidth]{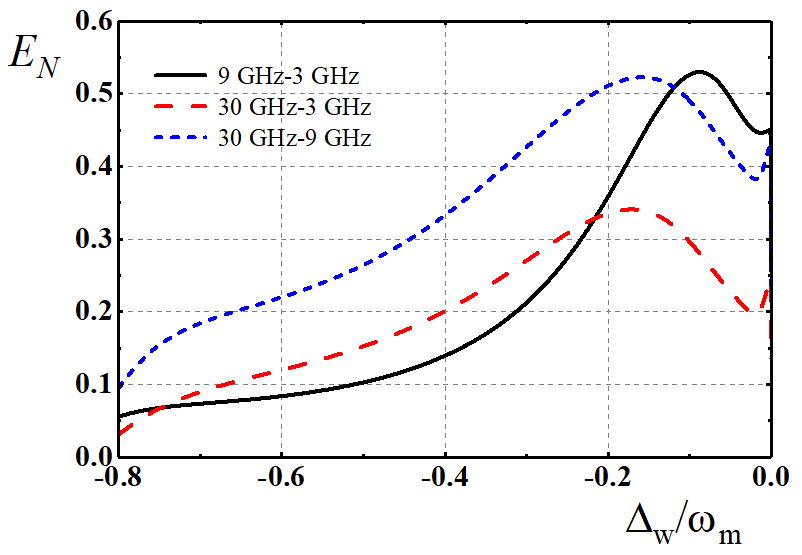}}
	\hspace{1in}
	\subfigure[ The microwave detuning range is from 0.1 to 0.8 $\Delta_{w}/\omega_m$.]{
		\label{fig:Fig2(b).png} 
		\includegraphics[width=1\linewidth]{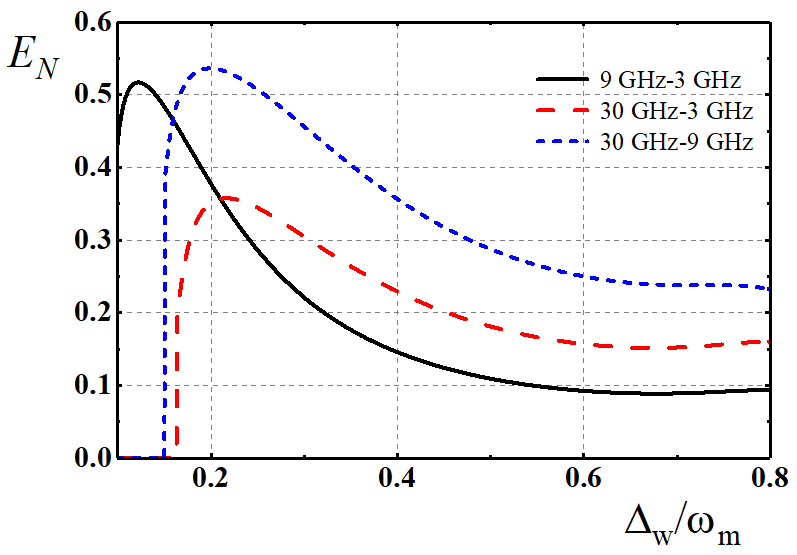}}
	\caption{Entanglement between two microwave with different frequency. (Full black line: $\omega_{w1}/2\pi$ = 9 GHz and $\omega_{w2}/2\pi$ = 3 GHz. Dashed red line: $\omega_{w1}/2\pi$ = 30 GHz and $\omega_{w2}/2\pi$ = 3 GHz. Dotted blue line: $\omega_{w1}/2\pi$ = 30 GHz and $\omega_{w2}/2\pi$ = 9 GHz.) The surrounding temperature T is fixed at 15 mK, the optical detuning $\Delta_c = \omega_m$, while the other parameters are $\lambda_{0c}$ = 1550 nm, $\kappa_c$ = 0.08 $\omega_m$, $P_c$ = 30 mW, $m$ = 10 ng, $\omega_m/2\pi$ = 10 MHz, $\kappa_{w1}$ = $\kappa_{w2}$ = 0.02 $\omega_m$, $P_{w1}$ = $P_{w2}$ = 30 mW, $d_1$ = $d_2$ = 100 nm, $\mu_1$ = $\mu_2$ = 0.008 and $\Delta_{w1}$ = $- \Delta_{w2}$$ \equiv\Delta_{w}$.}
	\label{fig:Fig2.png} 
\end{figure}
The simulation parameters are listed as follows. For the optical part, the driven laser wavelegth $\lambda_{0c}$ = 1550 nm, the damping rate $\kappa_c$ = 0.08 $\omega_m$, the driving power $P_c$ = 30 mW and the length of the cavity $L$ = 1 mm. For the mechanical part, the resonator mass $m$ = 10 ng, the resonate frequency ${{{\omega _m}} \mathord{\left/
		{\vphantom {{{\omega _m}} {2\pi }}} \right.
		\kern-\nulldelimiterspace} {2\pi }}$ = 10 MHz and its quality factor $Q$ = $5 \times {10^4}$. For microwave circuits part, we assume their damping rates $\kappa_{w1}$ = $\kappa_{w2}$ = 0.02 $\omega_m$, the input power $P_{w1}$ = $P_{w2}$ = 30 mW, the parameter related to the coupling $d_1$ = $d_2$ = 100 nm and $\mu_1$ = $\mu_2$ = 0.008. The frequency of the two microwave circuits and the optical detuning $\Delta_c$ will show up in the bottom of the Figures as well as the surrounding temperature. What's more, the microwave detunings are set to be opposite $\Delta_{w1}$ = $- \Delta_{w2}$ $ \equiv \Delta_{w}$.

Fig.2 shows the entanglement between two different frequency microwave modes. We chose three pairs of microwave frequencies for simulation, 9 GHz - 3 GHz, 30 GHz - 3 GHz, 30 GHz - 9 GHz. The biggest feature in Fig. 2 is that in an interval where the microwave detuning is larger than zero, the entanglement described by the three curves is abruptly going to zero, and for the beauty of the drawing, we blocked this part and divided it into two parts, respectively showing the characteristics of entanglement in the case where the microwave detuning is positive and negative. This is because the system does not satisfy the Routh-Hurwitz criterion in this interval, but the stationary and robust entanglement we discuss in this work is under the condition that the system is stable. If the system is unstable, the stationary and robust entanglement cannot be achieved. What's more, the width of these intervals varies with the frequencies of the entangled microwave pair, and the maximum entanglement is obtained on either side of the interval. There is a method to keep away from the adverse effects of the system instability caused by microwave tuning on the system, we chose the two microwave resonant frequencies to be the same. As we can see below, this can avoid the unstable situation and enable the system to be wideband tuned to adapt to a wider range application scenario.

The reason why we set the frequency of the two microwave cavities to be the same can avoid the instability of the system could be explained in the mathematical form. As we can see above, the parameters of the two microwave cavities are set to be the same except the resonant frequencies, and the two microwave detunings are the same value except a minus sign. If two microwave resonant frequencies are same, the two microwave circuits behave like twin circuits due to the structure parameters of two microwave circuits are completely the same. When we red-detune one microwave circuit, the other one is blue detuned and vice versa, which means under this condition the entanglement is an even function of the detuning, namely the entanglement on the red sideband is symmetrical to the one on the blue sideband. With this idea, we chose the microwave frequencies to be the same and show their entanglement properties in Fig.3.
\begin{figure}
\centering
\includegraphics[width=1\linewidth]{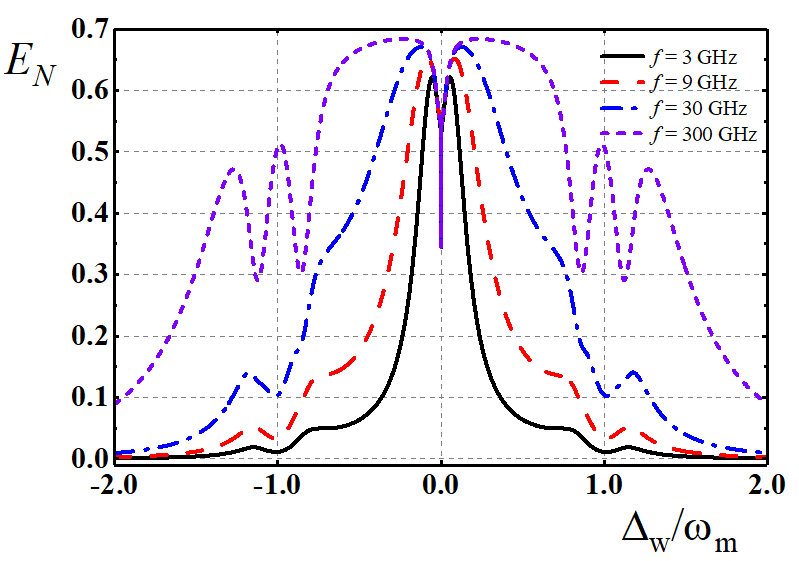}
\caption{Entanglement between two microwave with the same frequency. (Full black line: $\omega_{w1}/2\pi$ = $\omega_{w2}/2\pi$ = 3 GHz. Dashed red line: $\omega_{w1}/2\pi$ = $\omega_{w2}/2\pi$ = 9 GHz. Dotted dash blue line: $\omega_{w1}/2\pi$ = $\omega_{w2}/2\pi$ = 30 GHz. Dotted purple line: $\omega_{w1}/2\pi$ = $\omega_{w2}/2\pi$ = 300 GHz.) The other parameters are the same with Fig.2.}
\label{fig:Fig3.png}
\end{figure}

As shown in Fig.3, each curve is symmetrical based on the line of zero detuning, and the higher the frequency of the selected microwave pair, the larger the entanglement between the two microwave modes, and the maximum entanglement is obtained near the zero detuning. At this time, the system instability will not occur during the microwave broadband tuning process, that is, the Routh-Hurwitz criterion is always satisfied.
\begin{figure}
	\centering
	\includegraphics[width=1\linewidth]{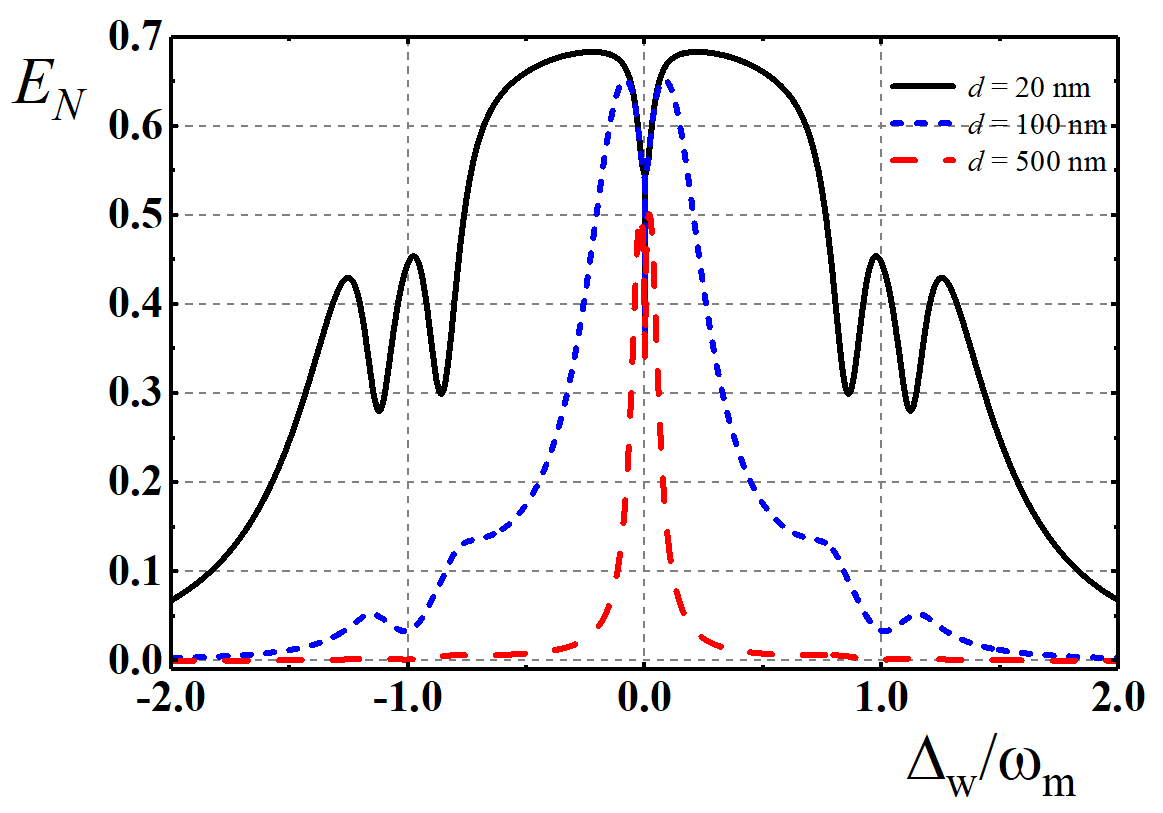}
	\caption{Entanglement between two microwave with the same frequency with different $d$. (Full black line: $d_1$ = $d_2$ = 20 nm. Dotted dash blue line: $d_1$ = $d_2$ = 100 nm. Dashed red line: $d_1$ = $d_2$ = 500 nm.) The microwave frequency $\omega_{w1}/2\pi$ = $\omega_{w2}/2\pi$ = 9 GHz. The other parameters are the same with Fig.2.}
	\label{fig:Fig4.png}
\end{figure}

Next we turn study the effect of architecture parameters of the hybrid system on microwave entanglement. In Fig.4, the smaller $d_1$($d_1$=$d_2$) contribute the better entanglement at the same microwave frequency $\omega_{w1}/2\pi$ = $\omega_{w2}/2\pi$ = 9 GHz. This could be explained by Eq.(35). The two microwave circuits are connected by the interaction with the optomechainical subsystem, and the electromechainical coulpings $G_{wj}$ increases with the $d_j$ decreases. When we increase the coupling $G_{wj}$ of two microwave circuits to the optomechainical subsystem, it also indirectly enhances the coupling between the two microwave circuits, which enlarges the entanglement between them.
\begin{figure}
	\centering
	\includegraphics[width=1\linewidth]{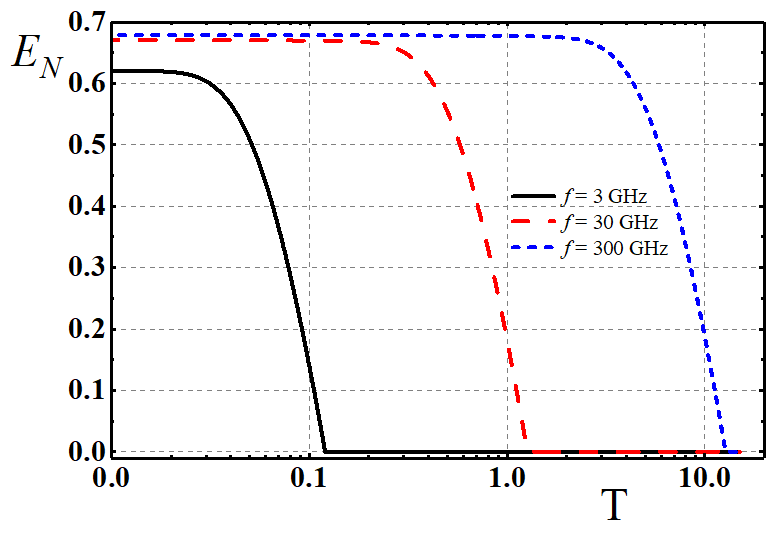}
	\caption{Entanglement between two microwave with different frequency for temperature. (Full black line: $\omega_{w1}/2\pi$ = $\omega_{w2}/2\pi$ = 3 GHz, $\Delta_{w}$ = $-$0.05$\omega_m$. Dashed red line: $\omega_{w1}/2\pi$ = $\omega_{w2}/2\pi$ = 30 GHz, $\Delta_{w}$ = $-$0.12$\omega_m$. Dotted blue line: $\omega_{w1}/2\pi$ = $\omega_{w2}/2\pi$ = 300 GHz, $\Delta_{w}$ = $-$0.13$\omega_m$.) The other parameters are the same with Fig.2.}
	\label{fig:Fig5.png}
\end{figure}

Fig.5 shows the entanglement of two microwave cavities versus the surrounding temperature at three pairs of different microwave frequencies: 3, 30, and 300 GHz, and the criterion of choosing the microwave detunings is to maximize entanglement. As the Fig.5 depicted, the higher the frequency of two microwave cavities, the better the temperature tolerance of entanglement between two microwaves. When the microwave frequency is selected to be 300 GHz, entanglement still exists above 10 K, this is because the microwave photon has higher energy and is more tenacious facing the thermal noise environment. 
\begin{figure}
	\centering
	\subfigure[Full plot, the microwave detuning range is from -2 to 2 $\Delta_{w}/\omega_m$.]{
		\label{fig:Fig6(a).png} 
		\includegraphics[width=1\linewidth]{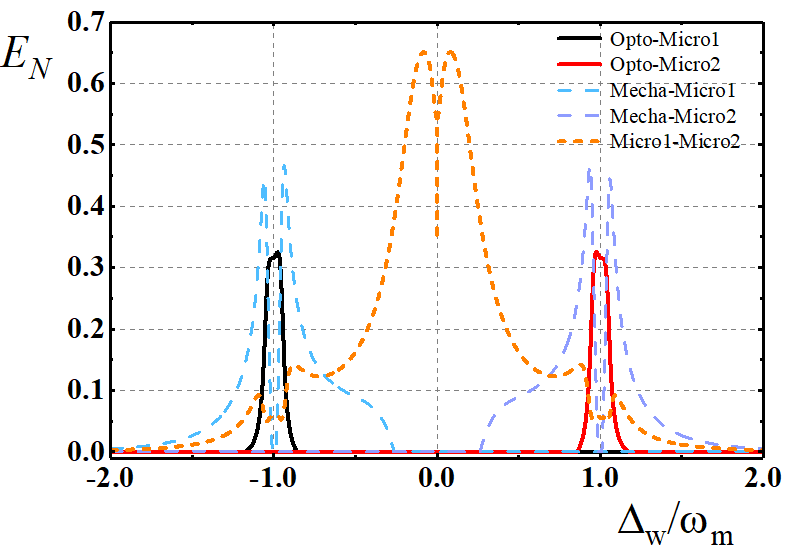}}
	\hspace{1in}
	\subfigure[Partial plot, the microwave detuning range is from -1.5 to -0.5 $\Delta_{w}/\omega_m$.]{
		\label{fig:Fig6(b).png} 
		\includegraphics[width=1\linewidth]{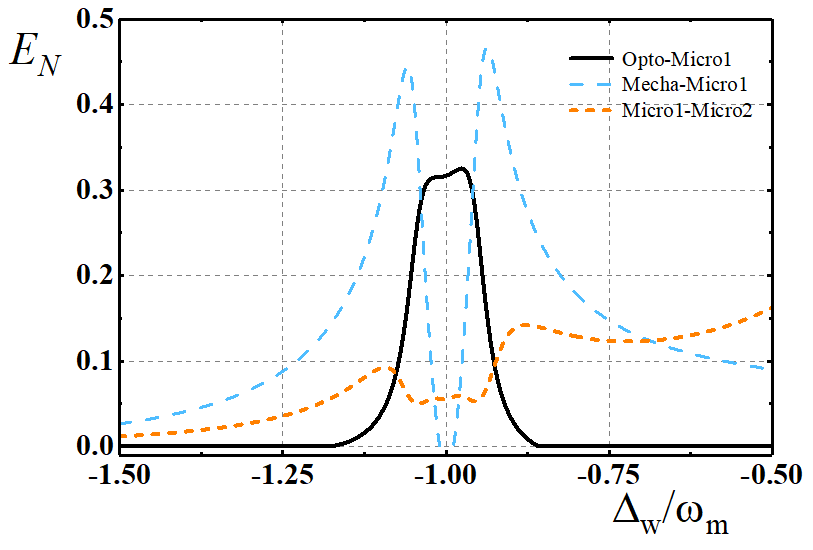}}
	\caption{The $E_N$ plot for five bipartite subsystems.(Full red and black line: two subsystems consist of optical cavity and one of the microwave circuits. Dashed blue and purple line: two subsystems consist of mechanical resonator and one of the microwave circuits. Dotted orange line: the subsystem consist of two microwave circuits. ``Opto'' means optical cavity, ``Micro1'' and ``Micro2'' mean the two microwave cavities and ``Mecha'' means the mechanical resonator) (a) is the full plot, and for the clarity, (b) is a part of (a). $\omega_{w1}/2\pi$ = $\omega_{w2}/2\pi$ = 9 GHz, $\kappa_c$ = 0.01 $\omega_m$, and the other parameters are the same with Fig.2.}
	\label{fig:Fig6.png} 
\end{figure}

Now we turn to study the entanglement between light and microwaves in this hybrid system. As shown in Fig. 6, it is known that the optical detuning $\Delta_c = \omega_m$, when $\Delta_c = \omega_m$ $ \approx $ $\Delta_{w}$, the entanglement between light and microwave reaches its maximum, while the entanglement between other subsystems, such as mechanical resonator-microwave circuits, microwave circuit and the other one, are compressed. This is because, in this case, the optical cavity and the microwave cavities are resonate through the optomechanical resonance and the electromechanical resonance, in other word, the entanglement between light and microwave mediated by mechanical resonator is at the expense of optomechanical and electromechanical entanglement.

In this section, we show the entanglement between two microwave circuits with different resonate frequencies, the influence of different parameters on the entanglement between two microwave modes with same resonate frequency and the entanglement of light and microwaves in the hybrid system. As shown in Fig.2, though the stationary and robust entanglement between two microwave circuits with different frequencies cannot be achieved in a small microwave detuning interval, we could make the microwave detuning controlled in a small interval, like from -0.4 to -0.1 on the $\Delta_{w}/\omega_m$ axis, and still obtain the stationary and robust entanglement, which has the potential to correlate different nodes in quantum computers, be the entangled microwave sources for quantum illumination radar or other applications. From Fig.3 to Fig.5, we can know that increasing the coupling of the two microwave cavities can acquire the larger entanglement between two microwave modes and raising the energy of microwaves can make the entanglement more tenacious in the thermal noise bath. Last but not least, Fig.6 shows the entanglement between microwaves and light is at the expense of optomechanical and electromechanical entanglement. In this way, we can obtain the entanglement of different bipartite subsystems in the hybrid quadripartite system under different microwave detuning by changing the detuning of the microwave circuit. For example, we set the microwave detuning $\Delta_{w}$ = $ \pm\omega_m$ and hybrid system now is used to generate the entanglement between light and microwave modes, furthermore, we set the microwave detuning $\Delta_{w}$ = $ \pm0.1\omega_m$ and hybrid system now is used to generate the entanglement between two microwave modes.

Now let us briefly discuss the promising experimental approach to realize the hybrid quadripartite system. As mentioned in reference [40], a vibrating $Si_3 N_4$ membrane coated in part with niobium interacts with an inductor–capacitor (LC) circuit that forms the microwave resonate cavity and couples by radiation pressure to an optically driven Fabry-Perot cavity. Our scheme could be an extension of the scheme mentioned above, the niobium could coat on the two sides of the membrane and interacts with two microwave circuits simutanouesly, which build the connection between two microwave circuits with optomechanics.

\section{Conclusion}
In this work, we theoretically proposed a scheme to build a hybrid system with the purpose of generate entanglement between two microwave modes, and the entanglement is measured by logarithmic negativity. The hybrid system consists of an Fabry-Perot cavity, a mechanical resonator, and two superconducting microwave circuits. It not only can generate the entanglement between the two microwave modes, but also can realize the entanglement of interested bipartite subsystems in the system, such as light and microwave modes, microwave modes, microwave mode and mechanical resonator and so on. Further, with different microwave detuning, the different bipartite subsystems of the hybrid system behave entangled. It means that we can choose proper microwave detuning to complete the requirements in the quantum information process, which microwave cavities are used to interface with solid-state qubits and light modes are used for quantum communication, or act as entangled microwave source to realize quantum illumination radar. Compared with previous theorectical work using logarithmic negativity as the measurement of entanglement [31,41-43], our maximum entanglment value $E_N$ is larger, which means we can in theory realize more entanglement than their schemes.

\begin{acknowledgments}
This work has been supported by National Key R$\& $D Program of China (2018YFA0307400); National Natural Science Foundation of China (NSFC) (61775025, 91836102).
\end{acknowledgments}
	
\bibliography{zhoupra}

\end{document}